\documentclass{optica-article}
\journal{opticajournal} 

\articletype{Research Article}
\usepackage[section]{placeins} 
\usepackage{subfigure}
\usepackage{multicol}
\usepackage{multirow}
\usepackage{array}
\usepackage{booktabs}
\usepackage{lineno}

\begin{document}

\title{Neural Network-Based Multimode Fiber Imaging and Characterization Under Thermal Perturbations}

\author{Kun~Wang,\authormark{1,$\dag$} 
Changyan~Zhu,\authormark{2,$\dag$} 
Ennio~Colicchia,\authormark{1}
Xingchen~Dong,\authormark{1}
Wolfgang~Kurz,\authormark{1}
Yosuke~Mizuno,\authormark{3,4}
Martin~Jakobi,\authormark{1}
Alexander~W.~Koch,\authormark{1}
and Yidong~Chong\authormark{2,*}}

\address{\authormark{1}Technical University of Munich, School of Computation, Information and Technology, Institute for Measurement Systems and Sensor Technology, Theresienstrasse 90, Munich 80333, Germany\\
\authormark{2}Division of Physics and Applied Physics, School of Physical and Mathematical Sciences, Nanyang Technological University, Singapore 637371, Singapore\\
\authormark{3}Faculty of Engineering, Yokohama National University, 79-5 Tokiwadai, Hodogaya-ku, Yokohama 240-8501, Japan\\
\authormark{4}Institute of Multidisciplinary Sciences, Yokohama National University, 79-5 Tokiwadai, Hodogaya-ku, Yokohama 240-8501, Japan\\
\authormark{$\dag$}The authors contributed equally to this work.}
\email{\authormark{*}yidong@ntu.edu.sg} 


\begin{abstract*} 
  Multimode fiber (MMF) imaging aided by machine learning holds promise for numerous applications, including medical endoscopy. A key challenge for this technology is the sensitivity of modal transmission characteristics to environmental perturbations. Here, we show experimentally that an MMF imaging scheme based on a neural network (NN) can achieve results that are significantly robust to thermal perturbations.  For example, natural images are successfully reconstructed as the MMF's temperature is varied by up to 50$^{\circ}$C relative to the training scenario, despite substantial variations in the speckle patterns caused by thermal changes.  A dense NN with a single hidden layer is found to outperform a convolutional NN suitable for standard computer vision tasks. In addition, we demonstrate that NN parameters can be used to understand the MMF properties by reconstructing the approximate transmission matrices, and we show that the image reconstruction accuracy is directly related to the temperature dependence of the MMF's transmission characteristics.
\end{abstract*}

\section{Introduction}

Fiber-optic imaging, in present-day commercial implementations, is predominantly performed using bundles of single-mode fibers (SMFs)~\cite{Gobel2004, Hughes2013}.  Such bundles, which consist of hundreds or even thousands of SMFs each transmitting one pixel of the target image, have been extensively and successfully employed in medical endoscopy and other applications.  However, there is a fundamental limitation to their performance: the image resolution is limited by the number of SMFs and the separation between the fiber cores, so increasing the resolution necessitates using thicker bundles and improved optics, thereby introducing other tradeoffs like decreased flexibility.

An alternative may be to use multimode fibers (MMFs), which have much larger core diameters than SMFs. Each MMF can simultaneously carry many thousands of propagating optical modes, allowing for extremely high information transmission capacity~\cite{cizmar2012}. A single MMF with a core diameter of 100~$\upmu$m supports approximately $10^4$ modes, which is a suitable number of information channels for many imaging applications~\cite{caramazza2019}. The difficulty, however, is that the modes are subject to mode-mixing and differing modal dispersion.  
This causes input images to be transformed into random-seeming patterns of speckles at the output, even though the actual light transmission in the MMF is fully deterministic~\cite{plschner2015, Beenakker1997}. Several ways of getting around this problem have been explored. In 1967, Spitz and Wertz demonstrated image transmission through an MMF using holography to correct for distortion, a precursor to what is now known as phase conjugation. This technique is limited by the requirement for the light to double-pass through the MMF, producing the final image only at the original object's location~\cite{spitz1967transmission}. 
Other methods, some originating in the related study of light propagation through turbid media~\cite{popoff2010, vellekoop2010, popoff2010NC}, have also been applied to MMF image reconstruction problem~\cite{Choi2012, Papadopoulos2013, Mahalati2012, NGom2018, Deng2018, Zhao2020}. However, these typically require complex instrumentation and extensive measurement or calibration efforts (e.g., to establish reference optical paths and determine the transmission matrix), rendering them impractical in many scenarios.

Another promising approach to MMF imaging is to employ machine learning algorithms based on neural networks (NNs). Aided by recent rapid improvements in hardware and software for running NNs, there have been several demonstrations of NN-based MMF imaging~\cite{Wang2018, Fan2021, Feng2022, Li_2020}. A major unresolved question is how well these methods can handle natural images under realistic environmental conditions; most of the demonstrations have been performed with simple images like the MNIST dataset, and using well-isolated fibers~\cite{Shabairou2018, Borhani2018, zhu2021, Ju2022, Song2022, Liu2022}. 
The reliability of the image reconstruction under environmental perturbations, such as thermal fluctuations, has yet to be systematically studied.

Here, we address these gaps by developing NN-based learning systems for MMF imaging and studying their performance under thermal perturbations. We investigate two different image datasets: MNIST (handwritten digits)~\cite{LeCun1998mnist} and ImageNet (natural images)~\cite{Jia_Deng_2009}. For both types of images, we are able to achieve accurate reconstruction across a wide temperature range of up to 50$^{\circ}$C difference between the training and testing sets. We test two NN architectures: a dense NN and a convolutional NN of the sort used in common computer vision applications.  
A previous study based on MNIST images had found that a dense NN performs better, apparently due to the nonlocality of the information encoded in the speckle patterns~\cite{zhu2021}. For natural images, we show here that the dense NN's advantage persists, but is somewhat reduced.  Finally, to help understand how the image reconstruction works, we develop a separate NN to learn an approximate transmission matrix (TM) of the MMF. Based on a singular value decomposition (SVD) mode analysis, we suggest that the former NNs achieve their results by learning certain fiber modes that are robust to speckle noise.

\section{Materials and methods}

\subsection{Experimental setup and data acquisition protocol}
\label{setup}

\begin{figure}
  \centering     
  \includegraphics[width=0.8\linewidth]{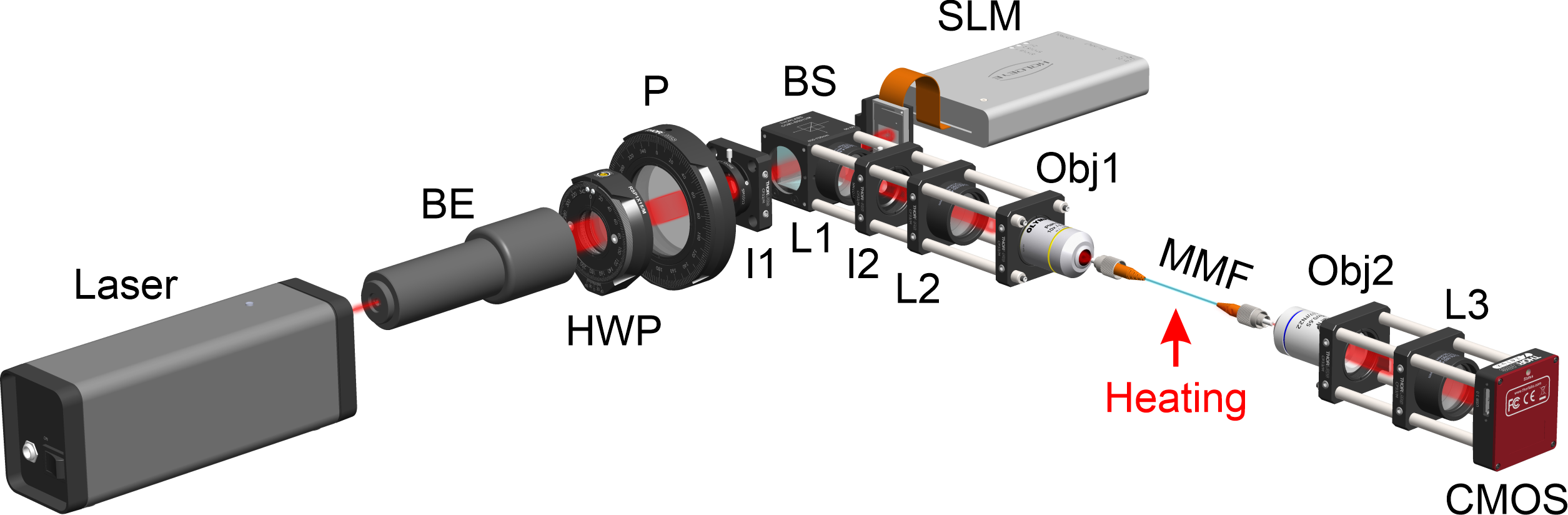}
  \caption{Schematic of the experimental setup: BE, beam expander; HWP, half-wave plate; P, polarizer; I1, iris; BS, beam splitter; SLM, spatial light modulator; L1, lens; I2, iris; L2, lens; Obj1, objective lens; MMF, multimode fiber; Obj2, objective lens; L3, lens; CMOS camera.}
  \label{fig.setup.imaging}
\end{figure}

\noindent The setup of our MMF imaging system is shown in Fig.~\ref{fig.setup.imaging}. A laser beam from a HeNe laser (HNLS008L-EC, Thorlabs), with wavelength 632.8~nm and power 0.8~mW, is expanded and collimated by a beam expander (BE) and filtered by a half-wave plate (HWP) and polarizer (P). The light is intensity-regulated and linearly polarized. The beam diameter is controlled by an iris (I1) and directed through a beam splitter (BS) onto a phase-only spatial light modulator (SLM, PLUTO-VIS, Holoeye).

The input pattern created by the SLM is filtered by another iris (I2) and then demagnified and imaged onto the entrance facet of the MMF through a 4\textit{$f$} imaging system composed of lens L1 and objective lens Obj1. The input images are drawn from either the MNIST dataset ($28\times28$ pixels each) or the ImageNet dataset ($64\times64$ pixels each). Each dataset pixel is aggregated into multiple SLM pixels, with the total SLM resolution being $1920\times1080$. More details about the image samples will be provided below.

The MMF has a core diameter of 400~$\upmu$m and a length of 45~cm. After the light passes through the fiber, the output speckle pattern is magnified and projected onto the CMOS camera by another 4\textit{$f$} system (lens L3 and objective lens Obj2). The resolution of the camera image is $1280\times1024$. To aid NN training, the collected speckle images are downsampled to $112\times112$ for MNIST and $128\times128$ for ImageNet.

A heating device induces thermal perturbations in the MMF during measurements. To separate the effects of the thermal perturbation from other uncontrolled perturbations that could also alter the MMF transmission characteristics, we acquire one set of speckles at a ``baseline temperature'' of 25$^{\circ}$C (run 1), and another set at a ``test temperature'' of 30$^{\circ}$C (run 2). Subsequently, the data collected at the baseline temperature (run 1) will be used to train the NN, and the data collected at the test temperature (run 2) will be used for testing. We repeat this procedure (including a fresh instance of run 1) for different values of the test temperature ranging from 30$^{\circ}$C up to 75$^{\circ}$C, as depicted in Fig.~S1 (Supplementary Material, Sec.~1).  The entire data acquisition process takes around 4 days.

\subsection{Neural network architecture and training}
\label{nnsetup}

For MMF image reconstruction, we consider two different NN architectures: a single-hidden-layer dense neural network (SHL-DNN) and U-Net~\cite{ronneberger2015u}. The latter is a convolutional NN commonly used in image recognition and related tasks. However, a previous study found that the simpler SHL-DNN architecture outperforms others in MMF reconstruction for simple input images, such as handwritten characters from the MNIST dataset~\cite{zhu2021}.

\begin{figure}
    \centering
    \vspace{-0.35cm}
    \subfigtopskip=0pt
    \subfigbottomskip=2pt
    \subfigcapskip=-6pt
	\subfigure[]{
		\begin{minipage}[t]{0.78\columnwidth}
		  \includegraphics[width=1\linewidth]{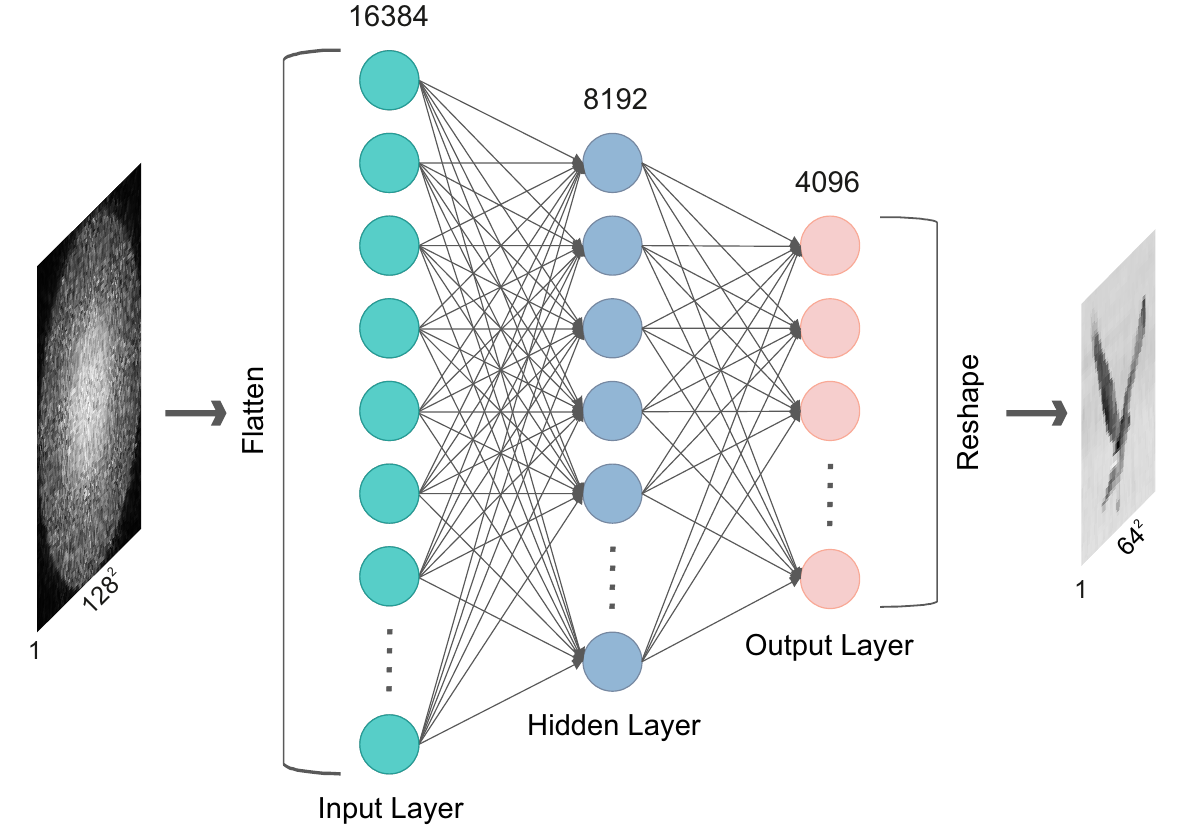}
		  \label{FCNN}
	\end{minipage}}
\\
	\subfigure[]{
		\begin{minipage}[t]{0.8\columnwidth}
			\includegraphics[width=1\linewidth]{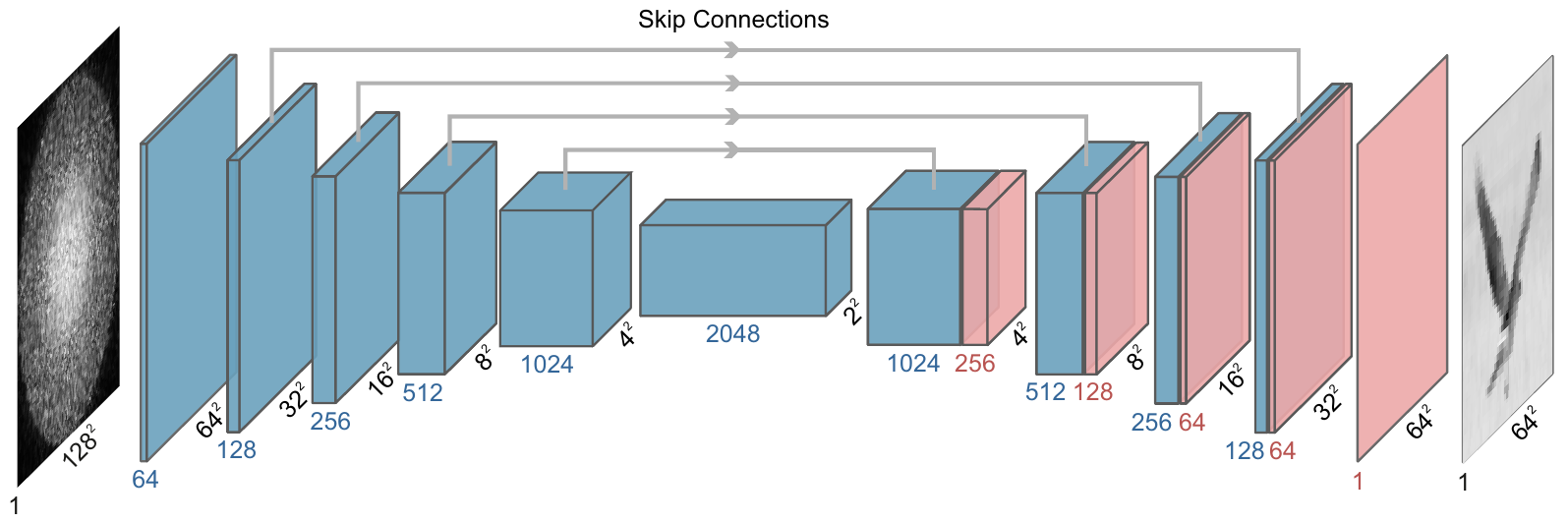}
			\label{Unet}
	\end{minipage}}
    \caption{Neural network architectures considered in this paper. (a) Single-Hidden-Layer Dense Neural Network, or SHL-DNN. (b) U-Net with skip connections. The input and output dimensions differ between MNIST and ImageNet images; the dimensions shown here are for MNIST images. }
    \label{network.architecture}
\end{figure}

The SHL-DNN, shown in Fig.~\ref{network.architecture}(a), comprises a single hidden layer with 8196 nodes, sandwiched between an input layer and an output layer, all fully connected. The U-Net is illustrated in Fig.~\ref{network.architecture}(b). It consists of 11 convolutional layers, each featuring a $3\times3$ kernel, beginning with the input layer for the speckle image. Following each convolutional layer in the encoder path is a nonlinear activation function (ReLu) and a max pooling operation with a stride of 2 for down-sampling. In the decoder path, we employ transposed convolutional layers as a substitute for max pooling layers to facilitate up-sampling. Both NN architectures are employed for ImageNet dataset training, whereas slight modifications are made for the MNIST dataset due to differences in image sizes, as detailed in the Supplementary Material.


A separate instance of the NNs is trained for each test temperature (the data collection scheme has been described in Sec.~\ref{setup}). In each instance, we have (i) 5000 pairs of ground truth and speckle images collected at the baseline temperature, randomly divided into a training dataset (4500 pairs) and a validation dataset (500 pairs); and (ii) 500 pairs collected at the test temperature, used as the test dataset. Batch normalization and dropout layers are used at each layer for both SHL-DNN and U-Net. The loss function is the mean square error (MSE).  
We use an Adam optimizer with learning rate decay, with the initial learning rate set as $10^{-6}$. The batch size is 2 with the number of epochs fixed at 50. All the training and testing procedures are conducted on the same computer (Nvidia DGX Station A100 Version 5.0.5, \texttt{GNU/Linux 5.4.0-80-generic x86\_64}).

\section{Results}

\subsection{MNIST dataset}
\label{sec:mnist}

First, we apply the MMF imaging scheme to $28\times28$ images of handwritten digits drawn from the MNIST dataset~\cite{LeCun1998mnist}. These images are substantially simpler than the natural images discussed in the next section. To quantify the performance of the NNs, we calculate the mean structural similarity index measure (SSIM)~\cite{wang2004image} over the set of 500 testing images at each test temperature.

\begin{figure}
    \centering     
    \includegraphics[width=\linewidth]{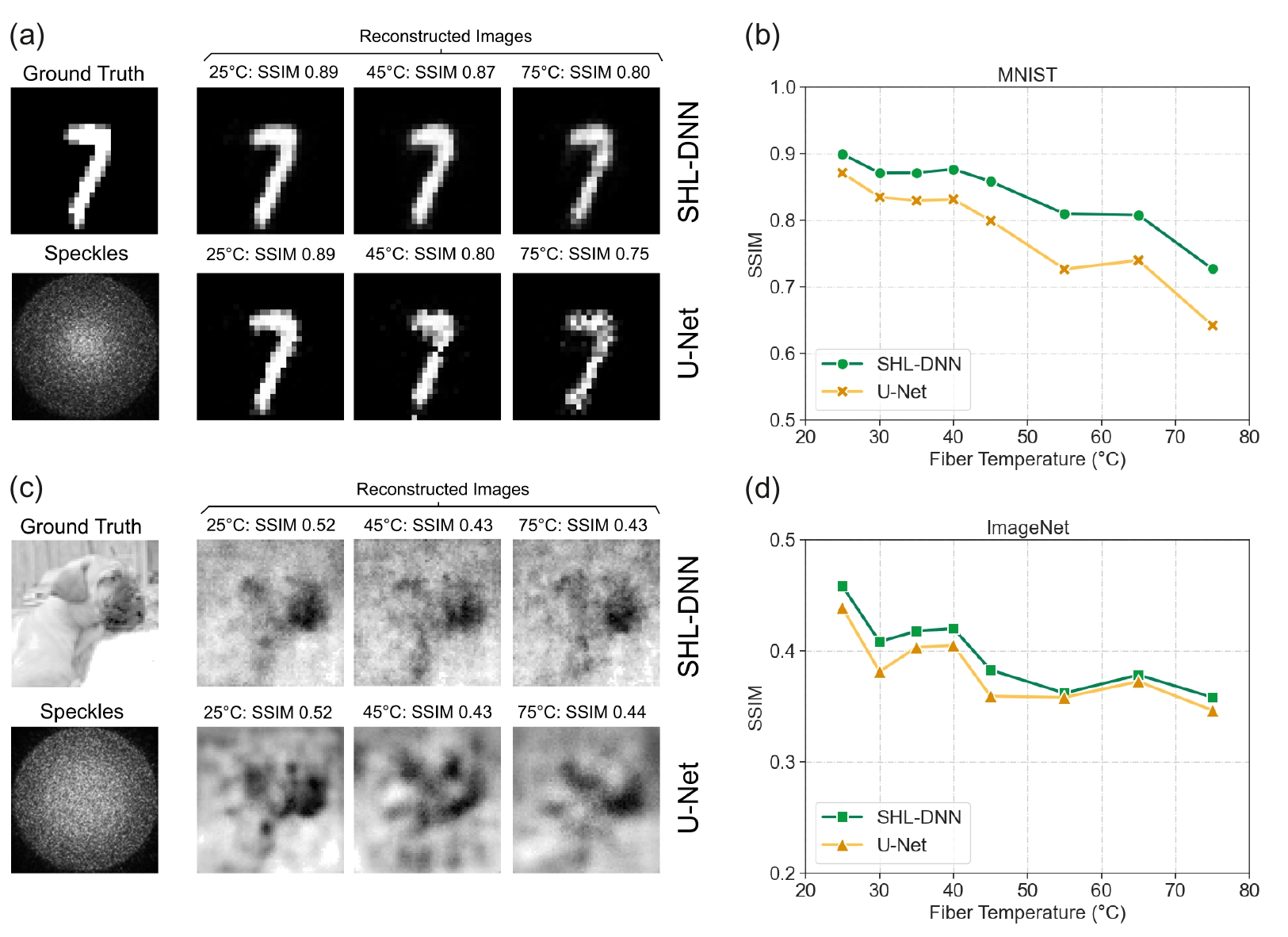}
    \caption{MMF imaging results.
    (a) A representative $28\times28$ ground truth image of the handwritten character ``7'' from the MNIST dataset; the speckle pattern after transmission through the MMF at 25$^{\circ}$C; and the images reconstructed by the two NNs when the fiber's test temperature is 25$^{\circ}$C, 45$^{\circ}$C, and 75$^{\circ}$C.  
    (b) Comparison of SSIM (a figure of merit for the image reconstruction accuracy, averaged over the test dataset) versus fiber temperature for MNIST images. Results are shown for the SHL-DNN and U-Net. 
    (c) A representative $64\times64$ natural image from the ImageNet dataset, and the speckle pattern and reconstructed images. 
    (d) The plot of the SSIM for the reconstructed natural images, averaged over the validation set, versus fiber temperature.}
    \label{fig.results}
\end{figure}

Both NNs reconstruct the images with high fidelity, maintaining legibility up to 75$^\circ$C, the highest tested fiber temperature.
Figure~\ref{fig.results}(a) shows results for a representative image from the test dataset, corresponding to the handwritten digit ``7''. In Fig.~\ref{fig.results}(b), we plot the temperature dependence of the mean SSIM, which declines by about 20\% (between 25$^\circ$C and 75$^\circ$C) for the SHL-DNN, and by about 27\% for the U-Net.  The better performance of the SHL-DNN is consistent with the findings of Ref.~\cite{zhu2021}.

\subsection{ImageNet dataset}

We then use the NNs to reconstruct natural images from ImageNet, a dataset designed for visual object recognition \cite{Jia_Deng_2009}. The input images in this case have a resolution of $64\times64$ and a speckle pattern resolution of $128\times128$. Therefore, for SHL-DNN, we change the hidden layer nodes to 4096, and for U-Net structure, we adjust the layer numbers to fit the new size (Supplementary Material, Sec.~2). The other training settings, such as the choice of loss function and optimizer, are kept the same.

Although previous studies have shown that NNs are able to reconstruct natural images through MMF speckles~\cite{caramazza2019, xu2023high}, these results were obtained using large training sets of around $5\times10^4$ samples. Our study uses an order of magnitude less data: $5\times10^3$ samples per training and validation dataset. At 25$^{\circ}$C (i.e.,~test temperature equal to the baseline temperature), we achieve a mean test SSIM of 0.45 for the SHL-DNN, and 0.43 for the U-Net.

When the thermal perturbation is applied, the image reconstruction for both NNs remains robust, even up to 75$^{\circ}$C. Fig.~\ref{fig.results}(c) showcases the results for three representative natural images reconstructed at different temperatures. For all test temperatures, the SHL-DNN achieves higher SSIM scores than the U-Net, as shown in Fig.~\ref{fig.results}(d), but its relative advantage is reduced compared to the MNIST task.


See the Supplementary Material, Sec.~3 and 4, for more examples of MNIST and ImageNet reconstruction results.

\subsection{Characterizing speckle variations}

\begin{figure}
    \centering     
    \includegraphics[width=1\linewidth]{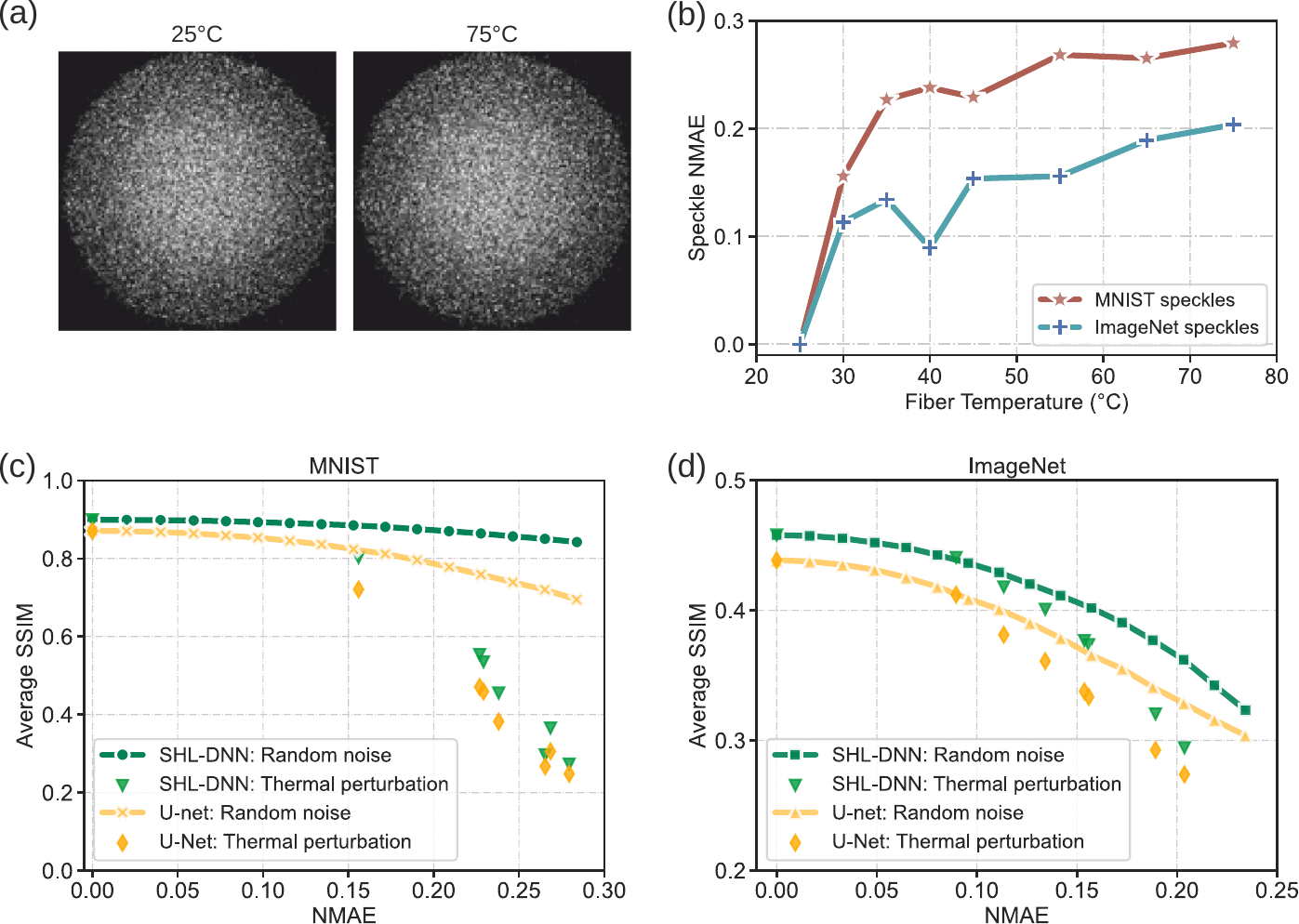}
    \caption{Effects of thermal variation on MMF speckle patterns and image reconstruction. (a) Two speckle patterns produced from the same input image at fiber temperatures of 25$^{\circ}$C and 75$^{\circ}$C.
    (b) Speckle pattern variations at different temperatures, quantified by their average NMAE values relative to the reference speckle pattern at 25$^{\circ}$C.
    (c,d) Average SSIM value of reconstructed images from two sets of speckle patterns. The first set consists of speckle patterns collected at different temperatures (thermal perturbations), while the second set comprises 25$^{\circ}$C speckle patterns with added random noises of varying intensity (resulting in different NMAE values). The reconstruction is performed using SHL-DNN and U-net trained on speckle patterns at 25$^{\circ}$C. (c) Results for the MNIST dataset. (d) Results for the ImageNet dataset.
    }
    \label{fig.recon_SSIM_speckle_NMAE}
\end{figure}


Figure~\ref{fig.recon_SSIM_speckle_NMAE}(a) shows two representative speckle patterns produced from the same image at $25^{\circ}$C and $75^{\circ}$C. The variations in speckle patterns are attributed to temperature-induced changes in the MMF’s geometry and refractive index,
which alter the mode-mixing and interference within the fiber. In other words, the speckle patterns are altered by changes to the transmission characteristics of the MMF, \textit{not} by the direct addition of noise to the output images.

Our systematic experimental data can help establish this important point. First, we quantify how the speckle patterns vary. For two $N$-pixel images $\mathbf{A} = \{A_1, \dots, A_N\}$ and $\mathbf{B} = \{B_1, \dots, B_N\}$, the normalized mean absolute error (NMAE)~\cite{George_Sideratos_2007, William_2012} is
\begin{equation}
  \textrm{NMAE}(\mathbf{A}, \mathbf{B}) = \frac{\textrm{MAE}(\mathbf{A},\mathbf{B})}{\textrm{mean}(B)},
\end{equation}
where the mean absolute error (MAE) is
\begin{equation}
  \textrm{MAE}(\mathbf{A}, \mathbf{B}) = \sum_i^N \frac{A_i - B_i}{N}.
\end{equation}
Compared to the mean squared error (MSE), the NMAE more accurately captures variations across different intensity scales. As shown in Fig.~\ref{fig.recon_SSIM_speckle_NMAE}(b), the NMAE exhibits an increasing trend with fiber temperature for speckles originating from both MNIST and ImageNet images.


Using this, we examine how the NN performance is affected by thermal perturbations, relative to a comparable level of synthetic noise in the speckle patterns. Figure~\ref{fig.recon_SSIM_speckle_NMAE}(c)--(d) shows the mean SSIM scores versus NMAE for two cases: (i) our standard experimental protocol, where the NMAE variation is due to temperature changes, and (ii) synthetic white noise of varying intensity levels, which are added to the speckle patterns collected at the baseline temperature. At each value of speckle NMAE, the NN performance is significantly lower in case (i), for both MNIST images [Fig.~\ref{fig.recon_SSIM_speckle_NMAE}(c)] and ImageNet images [Fig.~\ref{fig.recon_SSIM_speckle_NMAE}(d)].  Moreover, the disparity is greater for larger NMAE. These results show that compensating for thermal perturbations is qualitatively distinct from, and more challenging than, standard denoising.

\section{Neural network-based MMF characterization}

In this section, we introduce a different NN system designed not for reconstructing MMF images from output speckle patterns but to effectively replicate the physical behavior of the MMF. The trained parameters are meant to aid the qualitative analysis of the underlying MMF properties and their variations under thermal perturbations. 

The transmission characteristics of the MMF can be quantified using a pseudo-inverse transmission matrix, $\mathcal{T}$, defined by~\cite{Zhao_Tianrui_2020}
\begin{equation}
  I_{\rm in} = \mathcal{T}\,I_{\rm out},
  \label{Iinout}
\end{equation}
where $I_{\rm in}$ and $I_{\rm out}$ are the intensity profiles at the input and output facets, respectively.  This relation is approximate, since the input image's coherent features (i.e., phase information) are neglected~\cite{Zhao2020}. We focus on $\mathcal{T}$, not the ``forward'' transmission matrix, because the outcome of Eq.~\eqref{Iinout} can be directly compared to the NN-based image reconstruction scheme under analysis.

Measuring $\mathcal{T}$ accurately in real-time is known to be challenging because of its large size, and because it is sensitive to perturbations to the MMF~\cite{yoon2012experimental,rosa2020mode}. For our experiments on ImageNet images, $\mathcal{T}$ has dimensions $M \times N$, where $M = 64\times64 = 4096$ and $N = 128\times128 = 16384$.

\begin{figure}
  \centering     
  \includegraphics[width=0.95\linewidth]{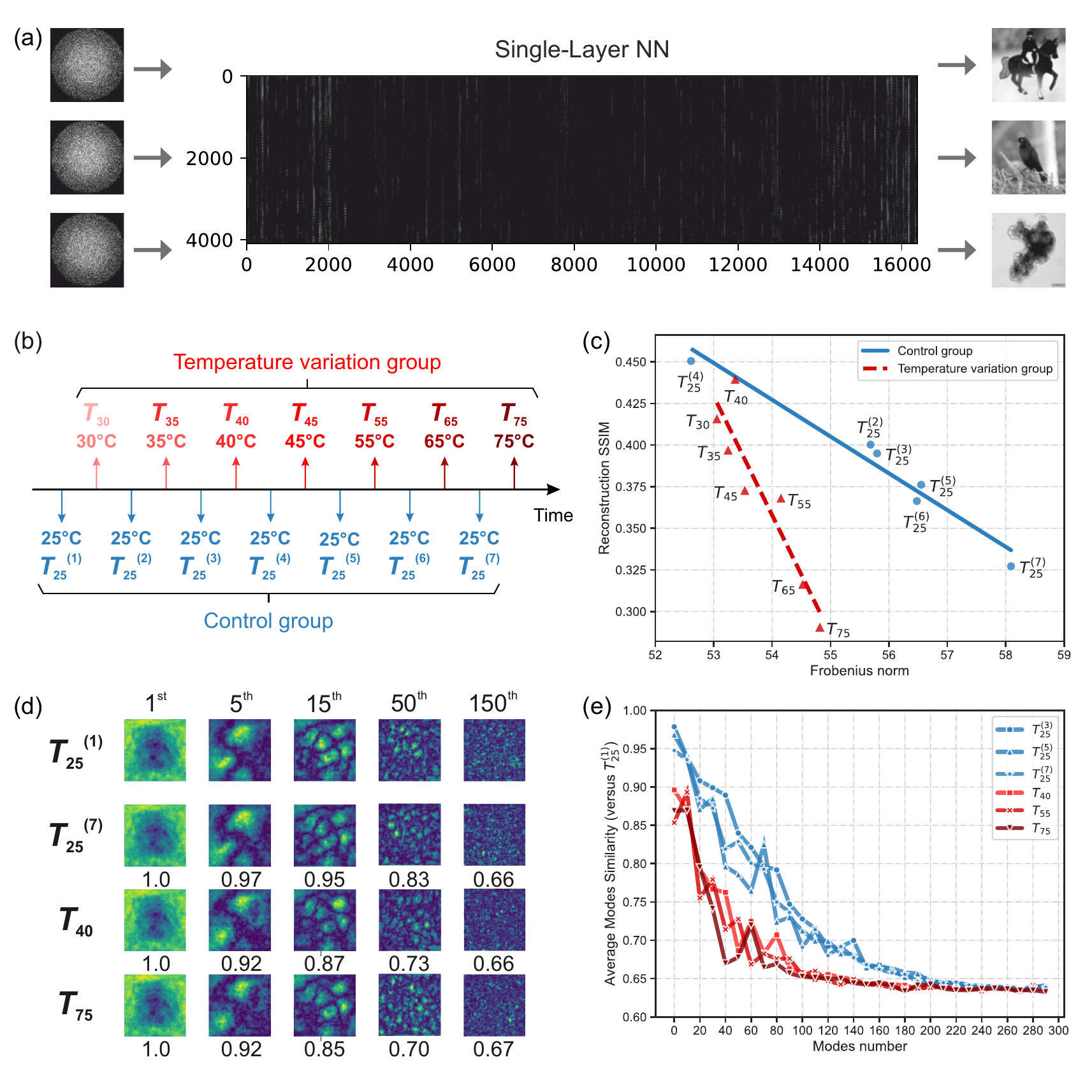}
  \caption{Analysis of MMF transmission characteristics. (a) Schematic of a single-layer NN trained with 500 pairs of speckle patterns and ground truth images. The NN parameters are then converted into an approximate inverse transmission matrix $\mathcal{T}$. (b) The datasets acquired at different temperatures and related $\mathcal{T}$s. 
  (c) Scatter plot of the mean image reconstruction SSIM (for different test temperatures) against the matrix Frobenius norm $\parallel \mathcal{T}_{\textrm{test}} - \mathcal{T}_{25} \!\parallel$ (where $\mathcal{T}_{25}$ and $\mathcal{T}_{\textrm{test}}$ are the corresponding approximate $\mathcal{T}$ matrices). Both the control group and temperature variation group show inverse proportional relationships, with Pearson correlation coefficients of -0.977 and -0.921, respectively. (d) Selected SVD modes for the control and temperature variation groups. The number below each image indicates the mode similarity, i.e., the inner product with the corresponding SVD mode for $\mathcal{T}^{(1)}_{25}$. (e) Average mode similarities, computed in the same way as in (d), versus mode number. Each point is an average over 10 SVD modes.}
  \label{fig.svd_modes}
\end{figure}

We now introduce an NN to serve as a workable approximation for $\mathcal{T}$.
This is a single-layer NN without an activation function (i.e., a real matrix), as shown in Fig.~\ref{fig.svd_modes}(a). For each test temperature, we train the single-layer NN using 500 pairs of images from the training dataset, using automatic differentiation to update the parameters.  Each set of $4096\times16384$ fitted NN parameters is then converted into an approximate $\mathcal{T}$ matrix. The resulting models are deliberately overfitted: they perform well on the training dataset, but poorly on the test dataset (see Supplementary Material, Sec.~5). This is acceptable because these models are intended for analyzing MMF characteristics, not for image reconstruction.


Using the datasets we had collected at different temperatures, we obtain a series of transmission matrices. As shown in Fig.~\ref{fig.svd_modes}(b), we divide them into two groups according to the temperature. The first group is the control group with only 25$^{\circ}$C datasets ($\mathcal{T}^{(1)}_{25}, \mathcal{T}^{(2)}_{25},...,\mathcal{T}^{(7)}_{25}$ ), and the second group is the temperature variation group ($\mathcal{T}_{30},...,\mathcal{T}_{75}$). Intuitively, the control group should capture how the MMF's transmission characteristics ``drift'' over time, while the temperature variation group includes both drifting and thermal effects.
We calculate the Frobenius norm of the difference between $\mathcal{T}^{(1)}_{25}$ and each of the other $\mathcal{T}$'s. In Fig.~\ref{fig.svd_modes}(c), we plot the Frobenius norm against the mean reconstruction SSIM for the same pairs of datasets.

We first consider the data for the temperature variation group (red triangle) in Fig.~\ref{fig.svd_modes}(c). The mean reconstruction SSIM decreases with the Frobenius norm---i.e., the less the $\mathcal{T}$ matrix changes, the more accurate the image reconstruction, as is to be expected.  We find a similar decreasing relation with the control group (blue dots). It is notable that the two sets of data points do not align, indicating that the reconstruction accuracy is affected by variations in the physical $\mathcal{T}$ matrices (due to drift and/or temperature change) that the Frobenius norm does not fully capture.

To gain more insight into the situation, we perform singular value decomposition (SVD) on the fitted $\mathcal{T}$ matrices (for more details, see Supplementary Material, Sec.~5). Each $\mathcal{T}$ matrix has size $M\times N$, and is decomposed as
\begin{equation}
  \mathcal{T} = U \, \Sigma \, V^{T}
\end{equation}
where $U$ and $V$ are real orthogonal matrices of size $M\times M$ and $N\times N$ respectively, and $\Sigma$ is a rectangular diagonal matrix of size $M\times N$. The column vectors of $U$, or ``SVD modes'', are eigenvectors of $\mathcal{T}\mathcal{T}^T$. We sort the SVD modes in decreasing order of their eigenvalues $|\sigma_i|^2$, which are the diagonal entries of $\Sigma \Sigma^{T}$  \cite{yoon2012experimental}.  

In Fig.~\ref{fig.svd_modes}(d), we present several representative SVD modes, reshaped to match the ground truth image dimensions. The number beneath each image indicates the overlap (inner product) with the corresponding $\mathcal{T}^{(1)}_{25}$ mode.

Figure~\ref{fig.svd_modes}(e) shows the similarity of modes across the first 300 modes, with each point representing the average over 10 modes. Overall, the mode similarity decreases as the mode number increases, reaching a floor of around 0.65 for higher-order modes. These results indicate that lower-order modes, which incorporate global and/or low-frequency spatial information, are more robust to environmental perturbations. Compared to the control group, the temperature variation group exhibits generally lower mode similarities. This is consistent with the findings of Fig.~\ref{fig.svd_modes}(c), and suggests that temperature variations induce a qualitatively different degree of variation in the MMF properties compared to the ordinary drift of the MMF's transmission characteristics, which leads to more serious degradation in image reconstruction accuracy.

\section{Conclusion}

In this study, we have systematically demonstrated NN-based multimode fiber imaging and characterization under thermal perturbations. Firstly, we show experimentally that two kinds of NNs---a single-hidden-layer dense NN and a convolutional NN (U-Net)---are effective at reconstructing an image inserted into the fiber from the speckle pattern collected at its end. This holds for both simple images (handwritten characters from MNIST) and natural images (from ImageNet). In each case, we find that (i) the reconstruction accuracy decreases with the difference in fiber temperature beween the training and test datasets, and (ii) the dense NN outperforms the convolutional NN, though the improvement is less pronounced for natural images.

In addition, we have demonstrated that NNs can be employed to analyze the transmission properties of MMFs. A dense NN trained on experimental data can approximate the MMF's transmission matrix, and variations in this matrix are well-correlated with observed performance degradations in the image reconstruction NNs. Using an SVD mode analysis, we find that lower-order modes are more robust, which helps explain why image reconstruction is possible across a wide temperature range. These findings enhance our understanding of MMF imaging under varying environmental conditions, with potential applications in various application domains where MMFs may experience significant temperature fluctuations. In the future, it would be desirable to extend the transmission matrix analysis to extract more information about the nature of wave propagation in this important class of random optical media.

\begin{backmatter}
\bmsection{Funding}
This work was supported by the Japan Society for the Promotion of Science (JSPS) KAKENHI (Grant No. 21H04555), and by the National Research Foundation (NRF), Singapore under Competitive Research Programme NRF-CRP29-2022-0003 and NRF Investigatorship NRF-NRFI08-2022-0001.


\bmsection{Disclosures}
The authors declare no conflicts of interest.

\bmsection{Data Availability}
Data underlying the results presented in this paper are not publicly available at this time but may be obtained from the authors upon reasonable request.


\bmsection{Supplemental document}
See Supplementary Material for supporting content. 

\end{backmatter}

\section{Appendix}


\bibliography{imaging}

\end{document}